\documentclass[aps,prl,floats,twocolumn,showpacs,superscriptaddress,floatfix]{revtex4}
\usepackage{dsfont}
\usepackage[dvips]{graphics}
\usepackage{graphicx}
\begin{document}

\title{Emergence of Cooperation in Non-scale-free Networks}

\author{Yichao Zhang}
\affiliation{Univ Normandy, France; ULH, LMAH, F-76600 Le Havre, FR CNRS 3335, ISCN, 25 rue Philippe Lebon, 76600 Le Havre, France}
\address{Department of Computer Science, University
College London, Gower Street, London, WC1E 6BT, United Kingdom}

\author{M. A. Aziz-Alaoui}
\email{aziz.alaoui@univ-lehavre.fr}
\affiliation{Univ Normandy, France; ULH, LMAH, F-76600 Le Havre, FR CNRS 3335, ISCN, 25 rue Philippe Lebon, 76600 Le Havre, France}

\author{Cyrille Bertelle}
\affiliation{Univ Normandy, France; ULH, LITIS, ISCN, F-76600 Le Havre, 25 rue Philippe Lebon, 76600 Le Havre, France}

\author{Shi Zhou}
\affiliation{Department of Computer Science, University
College London, Gower Street, London, WC1E 6BT, United Kingdom}

\author{Wenting Wang}
\affiliation{Department of Mathematics, University
College London, Gower Street, London, WC1E 6BT, United Kingdom}


\begin{abstract}
Evolutionary game theory is one of the key paradigms behind many scientific disciplines from science to engineering. Previous studies proposed a strategy updating mechanism, which successfully demonstrated that the scale-free network can provide a framework for the emergence of cooperation. Instead, individuals in random graphs and small-world networks do not favor cooperation under this updating rule. However, a recent empirical result shows the heterogeneous networks do not promote cooperation when humans play a Prisoner's Dilemma. In this paper, we propose a strategy updating rule with payoff memory. We observe that the random graphs and small-world networks can provide even better frameworks for cooperation than the scale-free networks in this scenario. Our observations suggest that the degree heterogeneity may be neither a sufficient condition nor a necessary condition for the widespread cooperation in complex networks. Also, the topological structures are not sufficed to determine the level of cooperation in complex networks.
\end{abstract}


\maketitle

\section{Introduction}
The long-term nature of competition leads to a confusing outcome that an agent's payoff does not depend on a decision in one round but on a reasonable updating rule~\cite{GTE,ETG,BAMS40479}. Thus in evolutionary game theory, the updating rule has played a key role, for instance, the well known Tit-for-Tat~\cite{TEC}, Pavlovian strategies~\cite{TD2647}, Win-Stay-Lose-Shift~\cite{Nature36456,TD35107}, Stochastic reactive strategies~\cite{GEB11364}, among others. Borrowing the concept of complex networks~\cite{RMP7447,AP511079}, people realized that abstracting interpersonal relationships into social networks~\cite{SIAMR45167,PR424175} may be an effective way to understand games in real human society.

To understand the pattern of games on social networks, a variety of network models were then intensively investigated~\cite{PR44697}, such as small-world networks~\cite{PRL95098104,PRE72056128} and scale-free networks~\cite{PRL95098104,PRL98108103}. In these networks, the group of opponents surrounding an individual is interpreted as its neighbors. The limited local interactions are the interactions restricted among the individual and their neighbors. As a successful attempt, the updating rule proposed by Santos and Pacheco~\cite{PRL95098104,PNAS1033490,JEB19726,NATURE454213} stood out. Simulation results~\cite{PRL95098104} show that the appearance of connected hubs in scale-free networks promotes the emergence of large-scale cooperation. This conclusion is striking in that it helps researchers find an origin of cooperative behavior in social networks. Also, it indicates that the interconnected hubs are usually occupied by cooperators, when selection favors cooperation. In Santos and Pacheco's updating rule, an individual randomly picks a neighbor as its reference. If the neighbor's payoff is higher than the individual, he/she adopt the neighbor's strategy with a certain probability. Given its feature of promoting cooperation in the networks with a power-law degree distribution~\cite{PRL95098104,PNAS1033490,JEB19726}, the rule is extensively employed in the following works~\cite{PRE72056128,PRL98108103,PNAS1033490,JEB19726,NATURE454213,EPL7730004}. Interestingly, their conclusion that degree-heterogeneity promotes the level of cooperation was then questioned theoretically in the public goods game~\cite{NJP13123027}. This difference indicates that the game model is a non-trivial factor as well. Indeed, the influence from updating rules or dynamics may be more influential~\cite{PLR6208,BS10282,BS10385}. Next, an empirical work reported by C. Gracia-L\'{a}zaro et al directly provides a realistic example~\cite{PNAS10912922}.

In this paper, we discuss an evolutionary game with a pure rational agent-based updating mechanism. In this game, we  consider a local deterministic nature selection. An individual always adopts a local better-performing strategy in the next round of game. Briefly, in a group including an individual and their neighbors, if the cooperative individuals can get a higher average payoff than the defective ones, the central individual will be a cooperator in the next round and vice versa. As in the well-mixed populations~\cite{GTE,EGPD}, for all the games investigated in this paper, defection dominates the population when the temptation to defect is powerful. However, for our updating rule, payoff memory~\cite{PRE88032127} can remarkably promote the levels of cooperation in the Watts and Strogatz's small-world network (WS)~\cite{NATURE393440} and random graphs, which clearly exceed that in the Barab\'{a}si and Albert's scale-free network (BA)~\cite{SCI286509}. Here, the payoff memory denotes the number of rounds during which the payoff of an individual is aggregated. Our observation indicates that degree heterogeneity may be neither a sufficient condition nor a necessary condition for the widespread cooperation in complex networks. In this respect, our conclusion is consistent with the recent empirical result~\cite{PNAS10912922}.

%
To stay consistent with the previous studies, we adopt the Prisoner's Dilemma (PD) as the game model. As a heuristic framework, the Prisoner's Dilemma describes a commonly identified paradigm in many real-world situations~\cite{EC,JCR,GCEC,NATURE398441}. It has been widely studied as a standard model for the confrontation between cooperative and selfish behaviors.
The selfish behavior here is manifested by a defective strategy, aspiring to obtain the greatest benefit from the interaction with others. This PD game model~\cite{ETG,GTE} considers two prisoners who are placed in separate cells. Each prisoner must decide to confess (defect) or keep silence (cooperate). A prisoner may receive one of the following four different payoffs depending on both its own strategy and the other prisoner's strategy.
\begin{itemize}
\item $T$ (temptation to defect) for defecting a cooperator.
\item $R$ (reward for mutual cooperation) for cooperating with a cooperator.
\item $P$ (punishment for mutual defection) for defecting a defector.
\item $S$ (sucker's payoff) for cooperating with a defector.

\end{itemize}
Normally, the four payoff values are defined as: $T>R>P\geq S$. At the next round, the prisoner will know the strategy of the other one in the previous round. It can then adjust its strategy according to the game updating rule.
For the case that two individuals just play the prisoner's dilemma for one round, there is a safe strategy, i.e., a prisoner $i$ always gets a payoff not less than his/her opponent if s/he defects. $i$'s strategy is denoted by $\Omega_i$. $\Omega_i$ takes vectors ${(1,0)}^T$ and ${(0,1)}^T$ for the cooperative and defective strategy respectively. For convenience, ${(1,0)}^T$ and ${(0,1)}^T$ are denoted by $\Omega_C$ and $\Omega_D$ hereafter. In one round of game playing with the other prisoner $j$, payoff $G_i$ can be rewritten as
\begin{equation}
G_i={\Omega^T_i}\left(
  \begin{array}{cc}
    R & S\\
    T & P\\
  \end{array}
\right){\Omega_j}.
\end{equation}

\section{Updating Rule}
\begin{figure}[hbtp]
\centering
\scalebox{0.4}[0.6]{\includegraphics[trim=0 30 0 0]{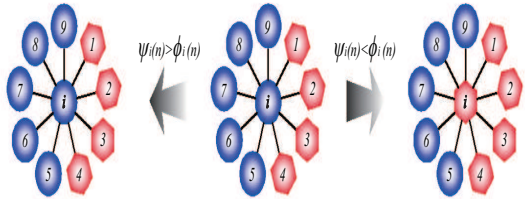}}
\caption{Illustration of strategy updating rule. For an individual $i$ with degree $9$, it has $5$ defectors (blue circles) and $4$ cooperators (red hexagons) as its neighbors in the $n_{th}$ round. The payoff of each neighbor is aggregated in the last $n-r\times \lambda$ rounds. $t$ is a value in the time region $\left[r\times \lambda+1,(r+1)\times \lambda\right]$. $G_x(n)$ denotes the aggregated payoff of neighbor $x$ ($x=1$, $2$, $...$, $9$), receiving from the $(r\times \lambda+1)_{th}$ round to the $n_{th}$ round. Based on Eqs.~\ref{Phi} and \ref{Psi}, ${\Phi}_i(n)=\frac{\sum_{t=r\times \lambda+1}^{n}G_1(t)+G_2(t)+G_3(t)+G_4(t)}{4}$ and ${\Psi}_i(n)=\frac{\sum_{t=r\times \lambda+1}^{n}G_5(t)+G_6(t)+G_7(t)+G_8(t)+G_9(t)+G_i(t))}{6}$.
}\label{update}
\end{figure}

In our updating rule, the PD is repeated in the following way: in the first stage, a PD is played by every pair of individuals connected by a link in the network. For an individual $i$ in the $n_{th}$ round, their payoff reads as
 \begin{equation}
G_i(n)=\sum_jA_{ij}\times{\Omega^T_i}(n)\left(
  \begin{array}{cc}
    R & S\\
    T & P\\
  \end{array}
\right){\Omega_j(n)},
\end{equation}
where $A_{ij}$ is an entry of the adjacency matrix of networks, taking values $A_{ij}=1$ ($i=1$, $2$, $...$, $the~size~of~networks$) whenever the individual $i$ and $j$ are connected and $A_{ij}=0$ otherwise. Each individual then updates their accumulated payoff, which is the sum of payoffs they receive from the last rounds in memory. The sum of payoffs for all the cooperative neighbors (defectors) of $i$ in the $n_{th}$ round is denoted by $C_i(n)$ ($D_i(n)$) as shown in Fig.~\ref{update}. We define $\lambda$ as the span of the payoff memory. For $\lambda=3$, $G_i(n)$ keeps aggregating for $3$ rounds. At the beginning of the $4$th round, the whole payoff system is reset. The purpose of introducing the payoff memory into the model is to simulate a points system of accumulating the players' payoffs. For example, a season in the English Premier League normally lasts $38$ rounds, in which $38$ denotes the memory span.
In the second stage, each individual updates their strategy based on their payoff and all their neighbors' payoffs. Consider a group of individuals including an individual $i$ and all $i$'s neighbors, if the cooperative individuals in this group have a higher average payoff than the defective ones, $i$ will be a cooperator in the next round, and vice versa. If the cooperative individuals' average payoff equals the defective ones' average payoff, $i$ will keep its strategy unchanged.
For the cooperative neighbors at the $n_{th}$ round, the sum of their payoffs reads as:
\begin{equation}
C_i(n)=\sum_j\sum_k A_{ij}\times \left(\Omega_{C}^T\cdot \Omega_{j}(n)\right)
\times\sum_{t=r\times \lambda+1}^nG_k(t),
\end{equation}
where $r\times \lambda+1 \leq t \leq (r+1)\times \lambda$ and $G_k(t)=\sum_jA_{jk}\times{\Omega_j}^T(n)\left(
  \begin{array}{cc}
    R & S\\
    T & P\\
  \end{array}
\right){\Omega_k(n)}$. The parameter $r=\lfloor \frac{n-1}{\lambda} \rfloor$. The function $\lfloor x \rfloor$ returns the largest integer lower than $x$. For the defective neighbors, the sum of their payoffs reads as:
\begin{equation}
D_i(n)=\sum_j\sum_k A_{ij}\times \left(\Omega_{D}^T\cdot \Omega_{j}(n)\right)
\times\sum_{t=r\times \lambda+1}^nG_k(t).
\end{equation}
Thus the average payoffs of the cooperative and defective neighbors are
\begin{equation}
{\Phi}_i(n)=\frac{{\left(\Omega_{C}^T\cdot {\Omega}_i(n)\right)}G_i(n)+C_i(n)}{{\left(\Omega_{C}^T\cdot {\Omega}_i(n)\right)}+\sum_j A_{ij}\times\left(\Omega_{C}^T\cdot {\Omega}_j(n)\right)}\label{Phi}
\end{equation}
and
\begin{equation}
{\Psi}_i(n)=\frac{{\left(\Omega_{D}^T\cdot {\Omega}_i(n)\right)}G_i(n)+D_i(n)}{{\left(\Omega_{D}^T\cdot {\Omega}_i(n)\right)}+\sum_j A_{ij}\times\left(\Omega_{D}^T\cdot {\Omega}_j(n)\right)},\label{Psi}
\end{equation}
respectively.

For $\Phi_i(n)=\Psi_i(n)$, $\Omega_i(n+1)=\Omega_i(n)$. For $\Phi_i(n)\neq\Psi_i(n)$, we derive
\begin{equation}
\Omega_i(n+1)=\left(\begin{array}{cc}
    \frac{|X(n)+1|}{2} \\
    \frac{|X(n)-1|}{2} \\
  \end{array}
\right),
\end{equation}
where
\begin{equation}
X(n)=\frac{\Phi_i(n)-\Psi_i(n)}{|\Phi_i(n)-\Psi_i(n)|}.
\end{equation}
The analytical process can be applied to other games by changing the payoff matrix $\left(
  \begin{array}{cc}
    R & S\\
    T & P\\
  \end{array}
\right)$. One can find out that the evolution process is deterministic after setting the initial roles.

\section{Gaming on Social Networks}

We consider four classes of networks, the WS small-world network~\cite{NATURE393440}, BA scale-free network~\cite{SCI286509}, regular and random graph. We generate ten WS networks, BA networks, regular graphs, and random graphs by random seeds. The regular graphs are formed by a number of individuals with an identical degree $6$. The WS networks and random graphs are generated by randomly rewiring $10\%$ and $100\%$ of the links, respectively. Each network has finite $1~024$ individuals and $3~072$ links. The BA networks are grown by attaching new individuals to $m=3$ existing individuals.

\begin{figure}[hbtp]
\centering
\scalebox{0.3}[0.3]{\includegraphics[trim=0 20 0 0]{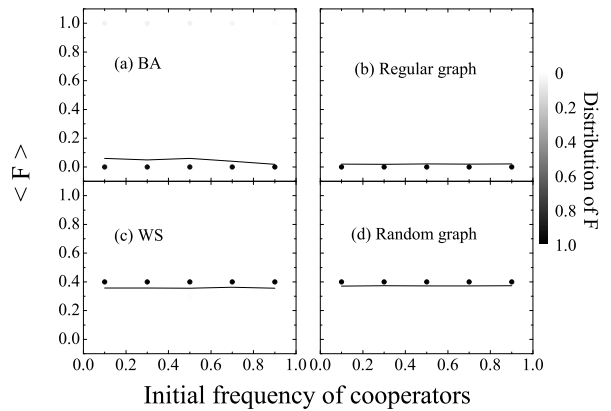}}
\caption{Average frequency $\langle{F}\rangle$ (solid lines), distribution of frequencies of cooperators (dots) as a function of initial frequency of cooperators for the Prisoner's Dilemma. We run 100 simulations for each of the parameter values for the game on each of the 10 networks. Thus each plot in the figure corresponds to 1~000 simulations with $\lambda=30$ and $b=1.9$. The colored circles are the binned data of $F$. For example, a decimal in $[0.25, 0.34)$ is approximated with $0.3$. $\langle{F}\rangle$ is obtained directly from the simulations. We consider four network models: ($a$) shows the simulation results obtained on the BA scale-free networks, which are generated by $m_0=m=3$~\cite{SCI286509}, where $m_0$ denotes the size of the initial fully connected network and $m$ denotes the number of links among a new node and the existing individuals in the network; ($b$) shows the simulation results obtained on the regular graphs, which are formed by 1~024 identical individuals of degree $6$; ($c$) shows the simulation results obtained on the WS small-world networks, which are generated by randomly rewiring 10\% of the links in the regular graphs; ($d$) shows the simulation results obtained on the random graphs, which are generated by randomly rewiring all the links~\cite{NATURE393440}.
}\label{initial}
\end{figure}

For the PD, the payoff parameters are set as $T = b$, $R = 1$, and $P = S = 0$, where $1< b\leq 2$ represents the temptation to defect~\cite{PRL95098104}. The larger $b$ is, the more favorable defection becomes. We denote $F$ as the frequency of cooperators after reaching a network gaming equilibrium. We run $51~000$ time steps for each simulation, in which $50,000$ steps to guarantee that the system reaches a dynamical equilibrium. Next, we measure and average $F$ from $50~000$ to $51~000$ steps, in which the standard deviation of $F$ is less than $0.1$. For a given network, we initially assign a fraction of individuals as cooperators at random, and the remaining individuals as defectors. The influences of the initial frequency of cooperators are shown in Fig.~\ref{initial}.

Fig.~\ref{initial} shows that the initial frequencies of cooperators have a weak connection with $\langle F\rangle$, which is the average frequency of cooperators in gaming equilibrium. For different initial frequencies of cooperators, after a period of initial turbulence, the system can always reach a dynamical equilibrium, where the number of cooperators (or defectors) is stabilized at the particular value with minimum fluctuation. Thus we set the initial frequency of cooperators to 0.5 in all the following simulations. In Fig.~\ref{initial}(c)(d), one can observe the WS small world networks and random graphs are clearly proper platforms of cooperation.

\begin{figure}[hbtp]
\centering
\scalebox{0.3}[0.3]{\includegraphics[trim=0 50 0 0]{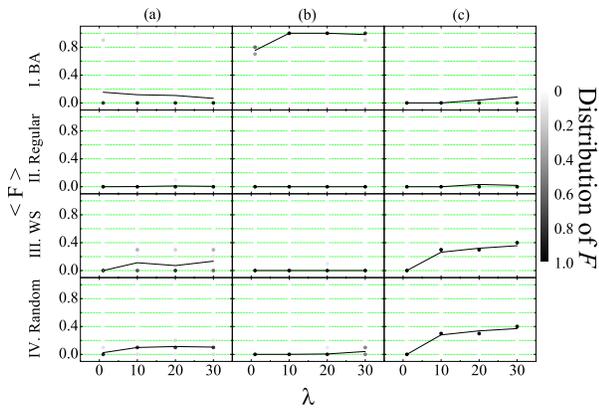}}
\caption{Average frequency $\langle{F}\rangle$ (solid lines), distribution of frequencies of cooperators (dots) as a function of $\lambda$ for the prisoner's dilemma with $b=1.9$. The column (a), (b), and (c) show the simulation results for the Nowak and May's updating rule~\cite{NATURE359826}, Santos and Pacheco's updating rule~\cite{PRL95098104}, and our updating rule, respectively. The initial frequency of cooperators is set to $0.5$. ({\bf I}), ({\bf II}), ({\bf III}), and ({\bf IV}) show the simulation results for BA scale-free networks, regular graphs, WS small-world networks, and random graphs, respectively.
}\label{game_mem}
\end{figure}
Fig.~\ref{game_mem} shows $\langle F\rangle$ as a function of the payoff memory span, $\lambda$, for the four types of networks.
The fraction of cooperators in a gaming equilibrium is determined by the network topology, game model, updating rule, payoff memory span and game parameters. The five factors govern the level of collaboration together. For $b=1.9$, panel I-(c) and II-(c) show that $\langle{F}\rangle$ is close to 0 in the BA scale-free networks and regular graphs. This observation indicates that collaboration is highly restrained in the system governed by our updating rule, even when the payoff memory is large. Instead, $\langle F\rangle$ can keep a relatively high level in the WS networks (see panel III-(c)) and random graphs (see panel IV-(c)) when the payoff memory is large. For the Santos and Pacheco's updating rule and the Nowak and May's updating rule, just few cooperators can survive in the regular and random graphs with a large payoff memory. The observations indicate that our model is a counter example of the previous conclusion~\cite{PRL98108103} that the scale-free networks provide a uniform platform of cooperation. Again, the behaviors observed in the WS small-world networks and random graphs indicate that even in a network without clear degree heterogeneity, cooperators still can survive in it.

\begin{figure}[hbtp]
\centering
\scalebox{0.3}[0.3]{\includegraphics{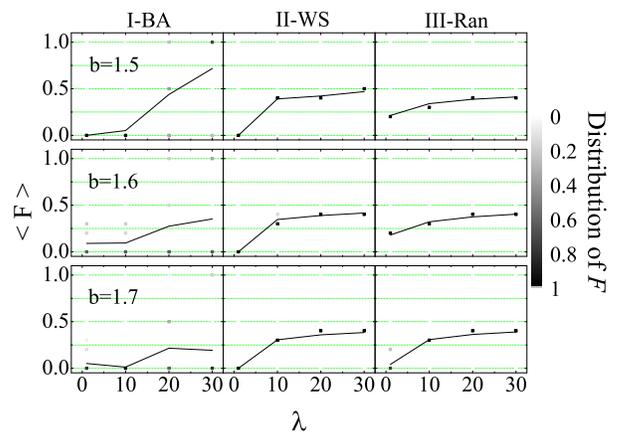}}
\caption{Average frequency $\langle{F}\rangle$ (solid lines), distribution of frequencies of cooperators (dots) as a function of $\lambda$ for the prisoner's dilemma with $b=1.5$ (the first row), $1.6$ (the second row), and $1.7$ (the third row), respectively. ({\bf I}), ({\bf II}), and ({\bf III}) show the simulation results for BA scale-free networks, WS small-world networks, and random graphs, respectively.
}\label{LA_1.5_1.6_1.7}
\end{figure}

Although we change both the updating rule and payoff memory to show the observations, these behaviors actually originate from the updating rule, since the payoff memory promotes cooperation in the social networks governed by the updating rules mentioned in this paper. For the Santos and Pacheco's updating rule with the payoff memory, the BA scale-free network is still the best platform of cooperation compared with the other topologies mentioned in this paper. In the BA scale-free network governed by our updating rule, the attraction of two extreme states $F=0$ and $1$ is much stronger than that in the WS small-world network and random graph. When the temptation to defect is powerful ($b=1.9$), the basin of attraction of full cooperation is smaller than full defection. However, in the WS small-world network and random graph, the attraction of full cooperation and defection does not exist. Thus the level of cooperation in the BA scale-free network is clearly lower than that in the WS small-world network and random graph. When the temptation to defect is not so large, for example $b=1.5$, the basin of attraction of full defection is not so large any more. Fig.~\ref{LA_1.5_1.6_1.7} shows that the payoff memory dramatically promotes the level of cooperation in the BA networks for $b=1.5$. In this case, the BA networks are still the proper platform of cooperation. For $b=1.6$ and $1.7$, the basin of attraction of full defection is large enough to restrain the level of cooperation in the BA networks. Thus the valid range of $b$ for our observations in Fig.~\ref{game_mem} is roughly $\left[1.6, 2.0\right]$.

\begin{figure}[hbtp]
\centering
\scalebox{0.3}[0.3]{\includegraphics{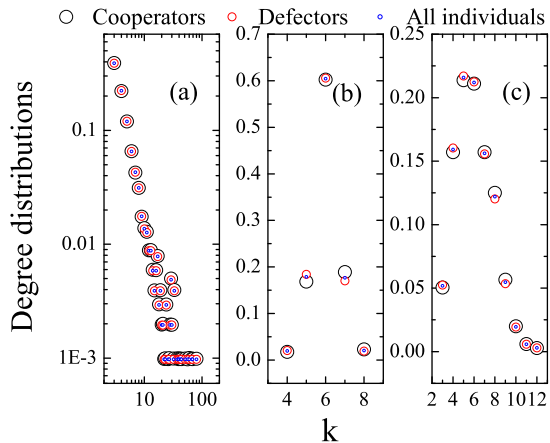}}
\caption{Degree distributions of cooperators and defectors in different topological structures. We consider (a) BA scale-free networks, (b) WS small-world networks, and (c) random graphs. For all the topological structures, we set $b=1.9$ and $\lambda=30$. Note that (a) is a double-logarithmic panel.
}\label{dis_norm}
\end{figure}

By comparing the BA network with the uncorrelated network (configuration model~\cite{RSA6161}), previous studies~\cite{PRL95098104} indicated that hubs are preferred by the cooperators. Interconnected hubs protect cooperation effectively~\cite{PRL95098104}. These two features whereas don't exist in our system. Fig.~\ref{dis_norm} shows the distributions of cooperators and defectors for BA networks, WS small-world networks, and random graphs, respectively. For these three classes of networks, one can observe that the distributions of cooperators are similar to the degree distributions of the networks. These observations indicate that cooperators do not preferentially occupy the hubs any more. They are evenly distributed.

\section{Discussion}
As shown in Fig.~\ref{game_mem} I-(b), Santos and Pacheco's updating rule~\cite{PRL95098104} dramatically promotes the level of cooperation in the BA scale-free networks, while it restrains cooperation in the other networks such as the WS small-world networks and random graphs. Based on an intuitive local strategy optimization, we introduce a new strategy updating rule to check whether the network structures are sufficient to determine the level of cooperation. Through extensive numerical simulations, we observe the network structure is just one of the key factors. In a proper condition, the level of cooperation in the WS small-world networks and random graphs can also exceed that in the BA scale-free networks.

To create such a proper condition, we consider the payoff memory in this paper. Actually, this parameter is not strange to most of us. In the previous game models, the payoff memory span used to be set to $1$, i.e., an individual's payoff in the current round will be cleared up after the strategy updating. In this condition, cooperation is highly restrained, even under our updating rule. With a small enhancement of the memory span, the evolutionary selection starts favoring cooperation. In term of this behavior, we presented a general explanation in our previous work~\cite{PRE88032127}. Admittedly, the influence of the payoff memory is limited. Nevertheless, it is sufficed to enable the random graphs and small-world networks to be proper platforms for the emergence of cooperation. We believe this example may shed some lights on the origins of the widespread altruistic behaviors in the non-scale-free networks.

\section{Conclusion}
In summary, we have discussed a question whether the degree heterogeneity promotes the level of cooperation for the Prisoner's Dilemma in complex networks.
Previous studies showed that the degree heterogeneity promotes the level of cooperation in the system governed by a particular strategy updating rule. A recent empirical study questions this conclusion in the case that humans play the Prisoner's Dilemma.
In this paper, we have proposed a strategy updating mechanism with payoff memory. Our simulation results show that the frequency of cooperators in the random graphs and WS small-world networks exceeds that in the BA scale-free networks apparently, especially when the temptation to defect is powerful.
Our observations indicate that the degree heterogeneity is neither a sufficient condition nor a necessary condition for the emergence of cooperation. In another word, the scale-free networks do not always promote cooperation in complex networks. In this respect, our results confirm the empirical result mentioned above. In a particular condition, cooperation can be widespread in any topological structure.
Our observations may provide a better understanding of the widespread cooperation in the networks without degree heterogeneity.

We wish to thank Martin Parsley and Wenxu Wang for their help with the manuscript. Y. Z. and M. A. A.-A. and C. B. are supported by the region Haute Normandie, France, and the ERDF RISC. S. Z. is supported by the UK Royal Academy of Engineering and the Engineering and Physical Sciences Research Council (EPSRC) under Grant No. 10216/70.

\end{document}